\documentclass[trackchanges]{aastex701}
\usepackage{multirow}
\usepackage{amsmath}
\usepackage{tikz}
\usepackage{booktabs}
\usepackage{float}
\usetikzlibrary{arrows.meta}
\usepackage{pgfplots}
\pgfplotsset{compat=1.18}

\begin{document}

\title{SMR: Scheduler with Multi-Channel Map-Encoded Reinforcement Learning for Radio Telescopes}

\author[orcid=0009-0009-8188-5632]{Zhenyang Huang}
\affiliation{
Xinjiang Astronomical Observatory, Chinese Academy of Sciences, Urumqi 830011, China
}
\affiliation{
School of Astronomy and Space Science, University of Chinese Academy of Sciences, Beijing 101408, China}
\email{huangzhenyang@xao.ac.cn}

\author[orcid=0000-0002-9786-8548]{Na Wang} 
\affiliation{
Xinjiang Astronomical Observatory, Chinese Academy of Sciences, Urumqi 830011, China
}
\affiliation{
Key Laboratory of Radio Astronomy and Technology (Chinese Academy of Sciences), A20 Datun Road, Chaoyang District, Beijing, 100101, P. R. China}
\affiliation{
Key Laboratory of Xinjiang Radio Astrophysics, Urumqi 830011, China
}
\email{na.wang@xao.ac.cn}

\author[orcid=0000-0002-2136-1708]{Zhiyong Liu} 
\affiliation{
Xinjiang Astronomical Observatory, Chinese Academy of Sciences, Urumqi 830011, China
}
\affiliation{
Key Laboratory of Radio Astronomy and Technology (Chinese Academy of Sciences), A20 Datun Road, Chaoyang District, Beijing, 100101, P. R. China}
\affiliation{
Key Laboratory of Xinjiang Radio Astrophysics, Urumqi 830011, China
}
\email{liuzhy@xao.ac.cn}

\author[orcid=0009-0006-1098-1174]{Chuhao Gao}
\affiliation{
Xinjiang Astronomical Observatory, Chinese Academy of Sciences, Urumqi 830011, China
}
\affiliation{
School of Astronomy and Space Science, University of Chinese Academy of Sciences, Beijing 101408, China}
\email{gaochuhao@xao.ac.cn}

\correspondingauthor{Na Wang, Zhiyong Liu}
\email{na.wang@xao.ac.cn, liuzhy@xao.ac.cn}

\begin{abstract}


Observation scheduling for large single-dish radio telescopes is a multi-objective optimization problem: schedulers must maximize on-source scientific return under strict mechanical and environmental constraints. Previous dynamic scheduling relies on expert-designed heuristics, while existing reinforcement-learning (RL) approaches often struggle with variable-length target lists and lack an intrinsic representation of sky geometry. We present SMR (Scheduler with Map-encoded Reinforcement Learning), which projects discrete targets onto an azimuth--elevation (Az--El) grid in the local horizon frame. The resulting aligned multi-channel sky maps encode target attributes together with direction-dependent cues such as satellite-interference risk and elevation-dependent receiver gain. This representation provides an explicit spatial inductive bias and enables SMR to learn directly from the sky state. 
Simulations based on real catalogs and site parameters show that, compared with a tuned look-ahead greedy baseline, SMR achieves about a 10\% relative improvement in time utilization by learning non-myopic scheduling strategies. In the full three-channel setting, SMR further achieves joint trade-off among efficiency, interference avoidance, and observation quality, with up to 17\% higher LIER and 54\% higher HGOR relative to an MLP baseline while maintaining higher utilization across both 12 h and 24 h horizons. Overall, SMR provides a simple and extensible way for data-driven single-dish scheduling.

\end{abstract}




\section{Introduction} \label{sec:intro}

Modern radio telescopes, especially large single-dish facilities, are central to a wide range of contemporary astronomical observations. 
With their high sensitivity and broad frequency coverage, they support a broad range of radio observations, including pulsar observations, spectral-line studies, continuum surveys, and time-domain radio astronomy. 
Modern radio observatories are highly integrated observing systems, combining multiple receivers, signal-processing backends, observing modes, and data acquisition pipelines \citep{fast,qtt,Rioja_2020}.

Representative facilities such as the Five-hundred-meter Aperture Spherical Radio Telescope \citep[FAST;][]{fast} and the under-construction Qitai Radio Telescope \citep[QTT;][]{qtt} require substantial investments in both construction and long-term operation. 
For such facilities, efficient use of observing time under operational constraints is therefore a central scheduling objective. 
In practice, this becomes a time-dependent scheduling problem: determining an observing sequence that balances source visibility, telescope slewing, instrument reconfiguration, and environmental constraints so as to maximize useful on-source observing time over a finite horizon. 
These constraints include weather and atmospheric opacity, source visibility in the local horizontal coordinate system, elevation limits and elevation-dependent performance, as well as avoidance constraints associated with bright sources such as the Sun.

Dynamic scheduling has therefore become a standard operational strategy at many observatories. 
For example, the Atacama Large Millimeter/submillimeter Array maintains a prioritized queue of Scheduling Blocks and selects an executable block according to real-time observing conditions \citep{ALMA_DynamicScheduling}. 
The Green Bank Telescope Dynamic Scheduling System uses an expert-designed scoring function that combines weather forecasts, observing band, instrument status, and project priority, and then optimizes schedules over a rolling 24--48 hour horizon \citep{ONeil2009GBT_DSS,Balser2009GBT_DSS_Algorithms}. 
These systems have been successful in routine operations, but they are typically built on expert-specified objectives and hand-tuned weights. 
Such designs can become difficult to extend when the constraint space is strongly nonlinear and when multiple operational trade-offs are tightly coupled.

Recent progress in deep reinforcement learning (RL), including Deep Q-Network (DQN), Proximal Policy Optimization (PPO), and Soft Actor--Critic (SAC), has demonstrated strong capability in sequential decision-making problems with high-dimensional state spaces \citep{DQN,PPO,SAC}. 
By interacting with a simulated environment, an RL agent can learn a policy directly from trial and error, reducing dependence on manually designed heuristics. 
This idea has already shown promise in astronomical scheduling. 
For example, \citet{nipsworkshop} showed that DQN could improve survey planning for the Stone Edge Observatory relative to traditional greedy strategies. 
More recently, the GRRIS framework represented a telescope array as a graph and combined Graph Neural Networks with RL to optimize observing strategies in time-domain astronomy \citep{GRRIS}.


Despite these advances, the application of modern RL methods to large radio telescope scheduling remains under-explored. 
First, radio telescope scheduling depends strongly on time-dependent source visibility and telescope pointing constraints. In the local horizontal coordinate system, the azimuth and elevation of each target evolve continuously with time, and elevation limits further make the set of visible targets change in size throughout an observing session. 
Many existing RL implementations handle this variable-length target list by flattening it into a fixed-length vector and passing it to a fully connected policy network. 
This representation provides little prior structure for the actual sky geometry and may generalize poorly when the number and spatial distribution of observable sources change.
A second challenge is that operational overheads are often a major part of the scheduling problem. 
Slewing between targets consumes time, switching receivers, observing modes or digital backends may introduce additional overheads, and azimuth cable wrap limits can eventually force costly unwrapping operations once the allowed rotation range is exhausted. 
These effects are especially important in pulsar observing programs, where individual integrations may be short but the target list can be large. 
In this regime, frequent slews and switches can dominate the observation efficiency, making it essential to account for these costs in the scheduler.

To address these challenges, we propose Scheduler with Map-encoded Reinforcement Learning (SMR), an end-to-end trainable framework with two key innovations. 
First, we represent the target list as an azimuth–elevation (Az–El) horizon map and process it with a convolutional network. 
This representation preserves local sky structure in the policy input and introduces a geometry-aware inductive bias that is more naturally matched to the telescope scheduling problem. 
Second, we adopt a two-stage training protocol. 
In the first stage, the agent learns the kinematic constraints associated with telescope motion, including slewing and azimuth cable wrap. 
In the second stage, we further introduce instrumental overheads. 
Together, these designs improve robustness and enable the scheduler to balance immediate observing reward against longer-term operational efficiency.

The remainder of this paper is organized as follows.
Section~\ref{sec:method} formulates the telescope scheduling problem and presents the proposed SMR framework, including the simulation setup, the map-encoded state representation, the action parameterization, the transition dynamics, the reward design, and the two-stage training protocol.
Section~\ref{sec:results} presents the main performance results across different scheduling horizons, analyzes the learned strategies under kinematic and switching constraints, and reports the key ablation studies.
Section~\ref{sec:multichannel_analysis} evaluates the scalability of the multi-channel Az--El map representation.
Section~\ref{sec:discussion} discusses the extensibility of SMR, its sensitivity, and its limitations in more realistic observatory settings.
Section~\ref{sec:conclusion} summarizes our main findings and outlines directions for future work.

\section{Method}\label{sec:method}

\subsection{Problem setup and notation}
We consider a single-dish radio telescope that executes a sequence of observations within a finite horizon
$\mathcal{H}=[t_0,t_0+H]$ discretized by a fixed step $\Delta t$ (in minutes).
Each day provides a candidate set of $N$ targets indexed by $i\in\{1,\dots,N\}$.
Each target $i$ is characterized by its equatorial coordinates $(\alpha_i,\delta_i)$, a required on-source integration time $d_i$ (minutes),
and an instrument configuration consisting of a receiver type $r_i\in\{1,2,3,4,5\}$ and a backend type $u_i\in\{1,2,3\}$.
Although receiver changes may be scheduled infrequently in practice, we treat $(r_i, u_i)$ more generally as discrete instrument configuration states that summarize any time-consuming overhead due to reconfiguration (e.g., setup, calibration, or observing-mode switching).

At each decision step, the scheduler selects the next target to observe.
Once a target $i$ is selected, the telescope executes an on-source observation of duration $d_i$ minutes,
and the completed target is removed from the remaining set.
We aim to build a scheduler that maximizes a proxy utility that combines on-source integration time and receiver gain, subject to realistic operational constraints, including sky visibility, slew kinematics, a unidirectional azimuth wrap limit, instrument switching overhead, and Sun avoidance.

\subsection{Observation catalog and simulation setup}\label{sec:data}
To emulate observatory operations, we generate day-scale scheduling instances using site-specific parameters and sky rotation.
We fix the telescope site at Urumqi, Xinjiang, China (latitude $44.03^\circ$\,N, longitude $89.58^\circ$\,E, altitude $781$\,m).
For each day instance, we construct a 24-hour time grid with 1-minute resolution and use \textsc{astropy} for coordinate transformations, converting the catalog ICRS coordinates into horizon coordinates $(Az_t^i, El_t^i)$ at each minute.
Mechanical visibility limits are $El\in[10^\circ,80^\circ]$ and $Az\in[0^\circ,360^\circ)$.

We define a binary visibility mask for each target:
\begin{equation}
W_i(t)=
\begin{cases}
1, & 10^\circ \le El_i(t) \le 80^\circ,\\
0, & \text{otherwise.}
\end{cases}
\end{equation}

We retrieve pulsars from the ATNF Pulsar Catalogue \citep{2005AJ....129.1993M} (VizieR: B/psr) and radio sources from the NVSS catalogue \citep{1998AJ....115.1693C} (VizieR: VIII/65/nvss) via the VizieR catalogue access service \citep{2000A&AS..143...23O}, using \textsc{astroquery} \citep{2019AJ....157...98G}.
Entries without valid equatorial coordinates $(\alpha,\delta)$ are discarded.
We then construct a daily target program by stratified sampling, with 40\% pulsars and 60\% NVSS sources.
For each selected target $i$, the required on-source integration time $d_i$ is drawn uniformly from $[1,10]$ minutes.
A target is retained only if its total visible time on a 1-minute grid satisfies $\mathrm{visible\ minutes} \ge d_i$.
We repeat the sampling procedure until the total requested on-source time $\sum_i d_i$ reaches $24\times60$ minutes.

To align the target catalogue with the map-encoded input, we discretize the local horizon into $1^{\circ}\times0.5^{\circ}$ cells in azimuth and elevation. The target map is therefore a finite-resolution representation of candidate availability. In the present implementation, each target-embedding pixel stores a single scalar value. When multiple targets fall into the same cell at the window start $t_0$, we randomly retain one representative target to keep the tensor representation single-valued at the adopted map resolution. This treatment does not assume a uniform sky distribution or impose any prescribed angular distribution on the catalogue. The non-uniformity of the target list remains represented by the spatial pattern of occupied cells.

For each retained target, we assign a receiver type $r_i$ uniformly from $\{1,2,3,4,5\}$ and a backend type $u_i$ uniformly from $\{1,2,3\}$, independently.
Each daily instance is stored as a CSV file in which each row corresponds to one target $i$.
The per-target fields are $\{\texttt{id}, r_i, u_i, \alpha_i, \delta_i, d_i\}$, where $\alpha_i$ and $\delta_i$
are the right ascension and declination (ICRS).
During simulation and policy execution, $(\alpha_i,\delta_i)$ are converted on the fly to the instantaneous horizon coordinates
$(Az_i(t), El_i(t))$ using the site location and current time.
$d_i$ is the requested on-source integration time (minutes).
Although training instances are generated with a 24-hour horizon, the learned scheduler is evaluated on arbitrary horizons $H$
(2, 4, 8, 12, and 24 hours).

We generate $1000$ training-day instances; each instance contains $N_{\rm day}\approx260$--$300$ targets, corresponding to a 24-hour budget.
For each training instance, the episode start time is set to the sampled window start $t_0$, and we initialize the telescope by selecting
one target uniformly at random as the first observed target.
For evaluation, we construct $20$ independent test instances for each horizon length $H\in\{2,4,8,12,24\}$ hours, resulting in $100$ test
instances in total.
Training instances follow the ``prefer-unused-sources'' sampling strategy described above, whereas test instances are randomly sampled under
the same feasibility constraints.
Although training uses $H=24$ hours, the learned scheduler is executed on an arbitrary observation window $\mathcal{H}=[t_0,t_0+H]$ at test time.

\subsection{State representation via map-encoded observation}\label{subsec:mapobs}
At each decision step $t$, the scheduler receives an observation that combines (i) the current telescope state and (ii) a horizon-aligned azimuth--elevation (Az--El) map that encodes the remaining targets.
This design allows the policy to directly exploit local sky geometry with a spatial inductive bias, while keeping the input dimension fixed even when the number of visible targets varies over time.

We represent the telescope pointing using sine--cosine features to avoid angular discontinuities:
\begin{equation}
\mathbf{s}^{\mathrm{tel}}_t
= \big[ \sin Az_t^{\mathrm{tel}},\ \cos Az_t^{\mathrm{tel}},\
        \sin El_t^{\mathrm{tel}},\ \cos El_t^{\mathrm{tel}},\
        r_t,\ u_t \big],
\end{equation}
where $Az_t^{\mathrm{tel}}$ and $El_t^{\mathrm{tel}}$ denote the current azimuth and elevation,
and $(r_t,u_t)$ denote the current receiver and backend types.
The discrete instrument indicators are normalized to $[0,1]$; all other non-angular scalars are min--max normalized.

We discretize elevation $El\in[10^\circ,80^\circ]$ with $0.5^\circ$ resolution and azimuth
$Az\in[0^\circ,360^\circ)$ with $1^\circ$ resolution, forming a $140\times360$ grid.
Let $(p,q)$ index the grid cell corresponding to a small Az--El patch.
At each time step $t$, we construct a three-channel tensor
\begin{equation}
\mathbf{M}_t\in\mathbb{R}^{140\times360\times3},
\end{equation}
by projecting each visible target $i$ to its instantaneous horizon coordinates $(Az_i(t),El_i(t))$ and writing
a channel-specific scalar into the corresponding cell.
In the present prototype, the target-embedding channel is constructed as a single-occupancy map following the collision-handling rule described in Section~\ref{sec:data}.
A detailed description of the three channels is provided below.

\paragraph{Channel 1 (target embedding).}
Each target is associated with target non-geometric attributes $(r_i,u_i,d_i)$ (receiver, backend, requested integration time).
We embed these attributes through a lightweight shared MLP into a scalar,
$e_\theta(r_i,u_i,d_i)\in\mathbb{R}$, and write it into the cell that contains target $i$.
Empty cells are set to zero.

\paragraph{Channel 2 (satellite-interference proxy).}
For the satellite-interference channel, we use the local angular density of propagated satellites on the Az--El grid as a time-dependent spatial penalty. The satellite population is constructed from the CelesTrak TLE sets for active satellites and stations, and the corresponding positions are propagated and projected onto the local horizon at each scheduling step. Cells containing more satellites are assigned larger penalty values, so that directions with higher satellite density are treated as less favorable for observation. This channel is intended as a proof-of-concept dynamic constraint for testing whether the SMR map-encoded interface can incorporate additional spatial optimization cues. It should not be interpreted as a predictive model of satellite radio-frequency interference (RFI) for a specific telescope, receiver, or frequency band. In operational use, the density-based proxy can be replaced by site- and band-specific RFI measurements, ephemeris-based avoidance masks, or receiver-safety constraints when saturation or hardware protection is relevant.



We obtain Two-Line Element (TLE) sets from CelesTrak\footnote{\url{https://celestrak.org/NORAD/documentation/tle-fmt.php}}, using the \texttt{active} and \texttt{stations} satellite groups. The two groups are merged, and duplicated entries are removed using their NORAD catalogue identifiers before propagation. We propagate the satellite orbits with the standard SGP4 model \citep{Vallado2006STR3,Vallado2008SGP4} (implemented with the \texttt{sgp4} package; \citealt{python_sgp4}). At each simulator step $t$ (with $\Delta t=1$ minute), we convert the SGP4 TEME state to apparent horizontal coordinates $(Az_s(t),El_s(t))$ for the observatory site using standard frame transformations as implemented in \textsc{astropy} \citep{astropy1,astropy2,astropy3}.

At each time step $t$, we construct a satellite-occupancy map on the Az--El grid and use it as a proxy for interference risk.
For each satellite, we first locate its central grid cell and assign it a value of 1.
To model the fact that interference can affect nearby pointings, we then ``spread'' each satellite to its local neighborhood with a simple discrete kernel:
cells one grid step away are assigned 0.9, two steps away 0.8, and so on, decreasing linearly to 0.1 at nine steps away.
Contributions from multiple satellites are accumulated, and the resulting map is finally rescaled to $[0,1]$ by min--max normalization,
yielding a normalized value $\lambda_s(t)\in[0,1]$ for every grid cell.
Under this convention, $\lambda_s(t)=0$ indicates no satellite influence in the cell, while larger values indicate denser (and thus potentially more interfering) satellite presence in that direction.

\begin{figure}[htbp]
  \centering
  \includegraphics[width=0.88\columnwidth]{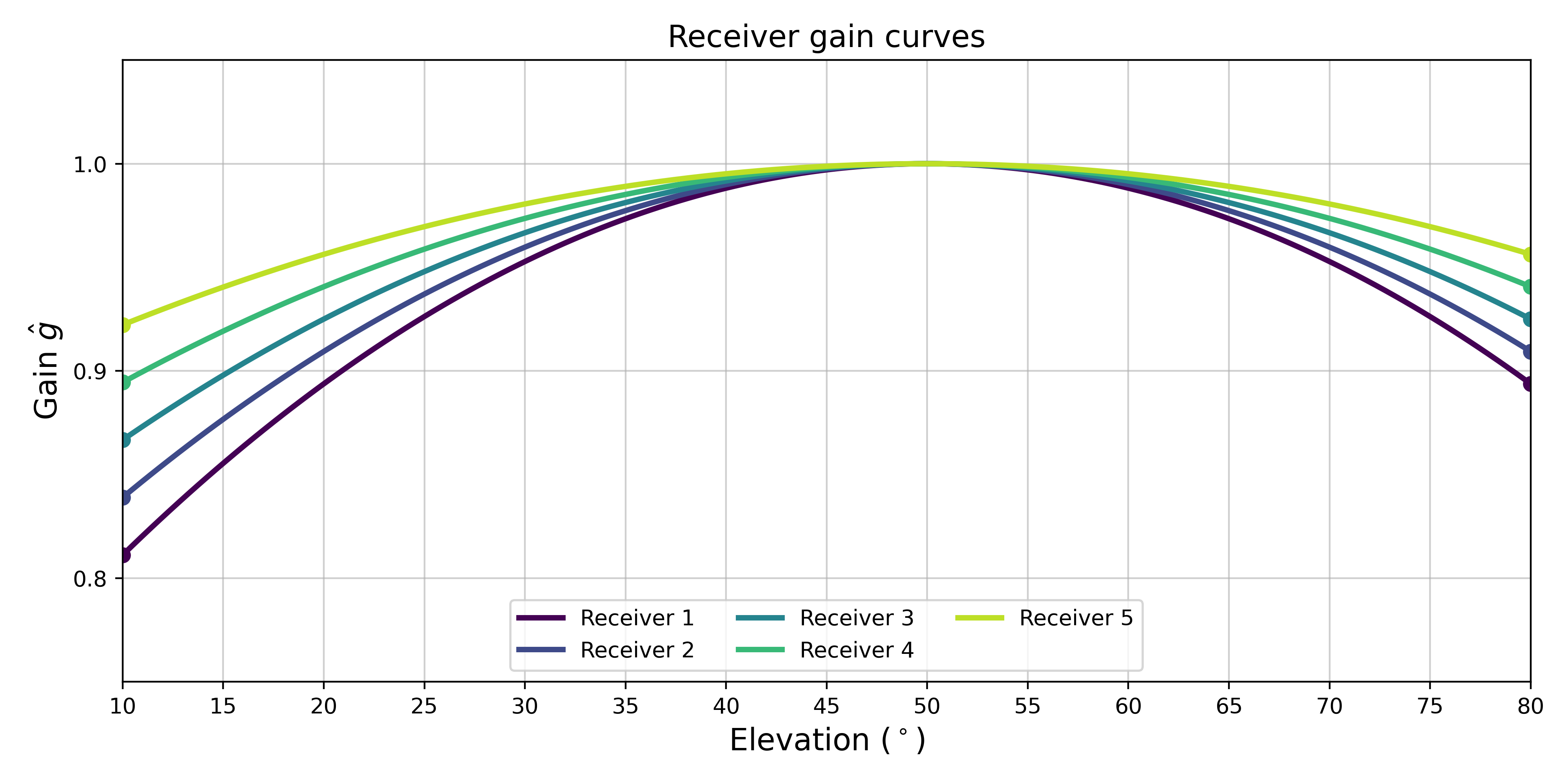}
  \caption{Illustration of receiver gain curves as a function of elevation.
  In SMR, this information is embedded into Channel~3 to encourage observing targets at optimal elevations.}
  \label{fig:gain}
\end{figure}

\paragraph{Channel 3 (receiver gain).}
For each target $i$, we query the normalized receiver gain at the current elevation,
$\hat g_{r_i}(El_i(t))\in[0,1]$, and write it into the corresponding cell.
Figure~\ref{fig:gain} illustrates the gain curves used in our simulator.
This channel encourages the policy to prefer elevations with higher system gain.

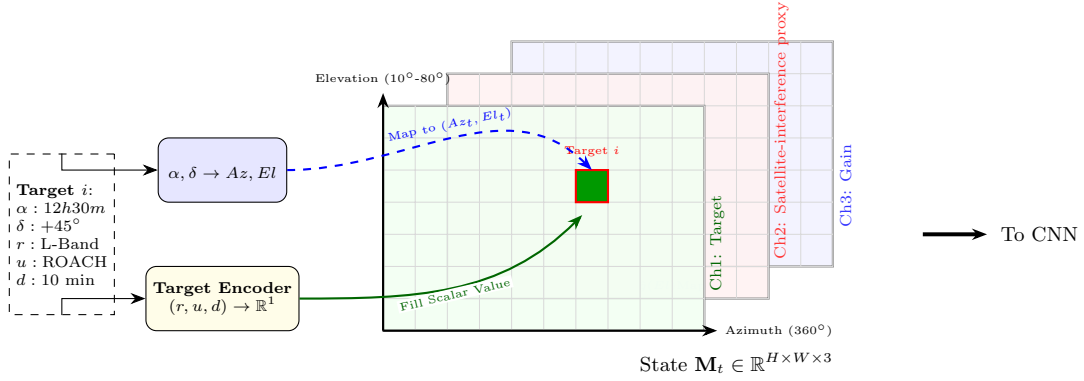
\begin{figure}[htbp]
    \centering
    \begin{tikzpicture}[
        scale=0.85,
        every node/.style={transform shape},
        >={Stealth[length=2mm]}
    ]
        \tikzstyle{process} = [rectangle, minimum width=2cm, minimum height=1cm, text centered, draw=black, fill=orange!10, rounded corners]
        \tikzstyle{data} = [rectangle, minimum width=1.5cm, minimum height=2.5cm, text centered, draw=black, dashed]
        \tikzstyle{layer} = [fill=white, draw=gray, opacity=0.9, thick]
        \tikzstyle{gridline} = [step=0.5cm, gray!30, very thin]

        \node[data, align=left, font=\scriptsize] (catalog) at (-5, 1.5) {
            \textbf{Target $i$}:\\
            $\alpha: 12h30m$\\
            $\delta: +45^{\circ}$\\
            $r: \text{L-Band}$\\
            $u: \text{ROACH}$\\
            $d: 10 \text{ min}$
        };

        \node[process, fill=blue!10, font=\scriptsize, align=center] (coord) at (-2.5, 2.5) {\\$\alpha, \delta \to Az, El$};
        \draw[->] (catalog.north) |- (coord.west);

        \node[process, fill=yellow!10, font=\scriptsize, align=center] (mlp) at (-2.5, 0.5) {\textbf{Target Encoder}\\$(r, u, d) \to \mathbb{R}^1$};
        \draw[->] (catalog.south) |- (mlp.west);

        \begin{scope}[shift={(2.0, 1.0)}]
            \draw[layer, fill=blue!5] (0,0) rectangle (5,3.5);
            \draw[gridline] (0,0) grid (5,3.5);
            \node[anchor=west, font=\scriptsize, text=blue!80, rotate=90] at (5.2, 0.5) {Ch3: Gain};
            \node[font=\tiny, text=blue!80] at (2.5, -0.3) {$\hat g(El)$Map};
        \end{scope}

        \begin{scope}[shift={(1.0, 0.5)}]
            \draw[layer, fill=red!5] (0,0) rectangle (5,3.5);
            \draw[gridline] (0,0) grid (5,3.5);
            \node[anchor=west, font=\scriptsize, text=red!80, rotate=90] at (5.2, 0.5) {Ch2: Satellite-interference proxy};
        \end{scope}

        \begin{scope}[shift={(0, 0)}]
            \draw[layer, fill=green!5] (0,0) rectangle (5,3.5);
            \draw[gridline] (0,0) grid (5,3.5);
            \node[anchor=west, font=\scriptsize, text=green!40!black, rotate=90] at (5.2, 0.5) {Ch1: Target};

            \draw[->, thick] (0,0) -- (5.2, 0) node[right, font=\tiny] {Azimuth ($360^{\circ}$)};
            \draw[->, thick] (0,0) -- (0, 3.7) node[above, font=\tiny] {Elevation ($10^{\circ}$-$80^{\circ}$)};

            \fill[green!60!black] (3.0, 2.0) rectangle (3.5, 2.5);
            \draw[red, thick] (3.0, 2.0) rectangle (3.5, 2.5);
            \node[font=\tiny, text=red] at (3.25, 2.8) {Target $i$};
        \end{scope}

        \draw[->, dashed, blue, thick] (coord.east) to[out=0, in=135]
            node[midway, above, font=\tiny, sloped] {Map to $(Az_t, El_t)$}
            (3.25, 2.5);

        \draw[->, thick, green!40!black] (mlp.east) to[out=0, in=225]
            node[midway, below, font=\tiny, sloped, fill=white, inner sep=1pt] {Fill Scalar Value}
            (3.1, 1.8);

        \node[right, font=\small] at (9.5, 1.5) {To CNN};
        \draw[->, line width=1.5pt] (8.4, 1.5) -- (9.4, 1.5);
        \node[font=\small] at (5.5, -0.5) {State $\mathbf{M}_t \in \mathbb{R}^{H \times W \times 3}$};
    \end{tikzpicture}
    \caption{Schematic of the Az--El map. Discrete target attributes are transformed and projected onto an azimuth--elevation grid to form a multi-channel spatial tensor that is ingested by the CNN-based policy network.}
    \label{fig:map_encoder}
\end{figure}

Figure~\ref{fig:map_encoder} provides a schematic overview of how discrete target attributes are projected to the Az--El grid.

\subsubsection{Final observation}
We use a CNN to encode $\mathbf{M}_t$ into a compact feature vector $\mathbf{m}_t$.
The final observation fed to the policy is
\begin{equation}
\mathbf{o}_t=\big[\,\mathbf{m}_t;\ \mathbf{s}^{\mathrm{tel}}_t\,\big].
\end{equation}

\begin{figure}[htbp]
  \centering
  \includegraphics[width=0.88\textwidth]{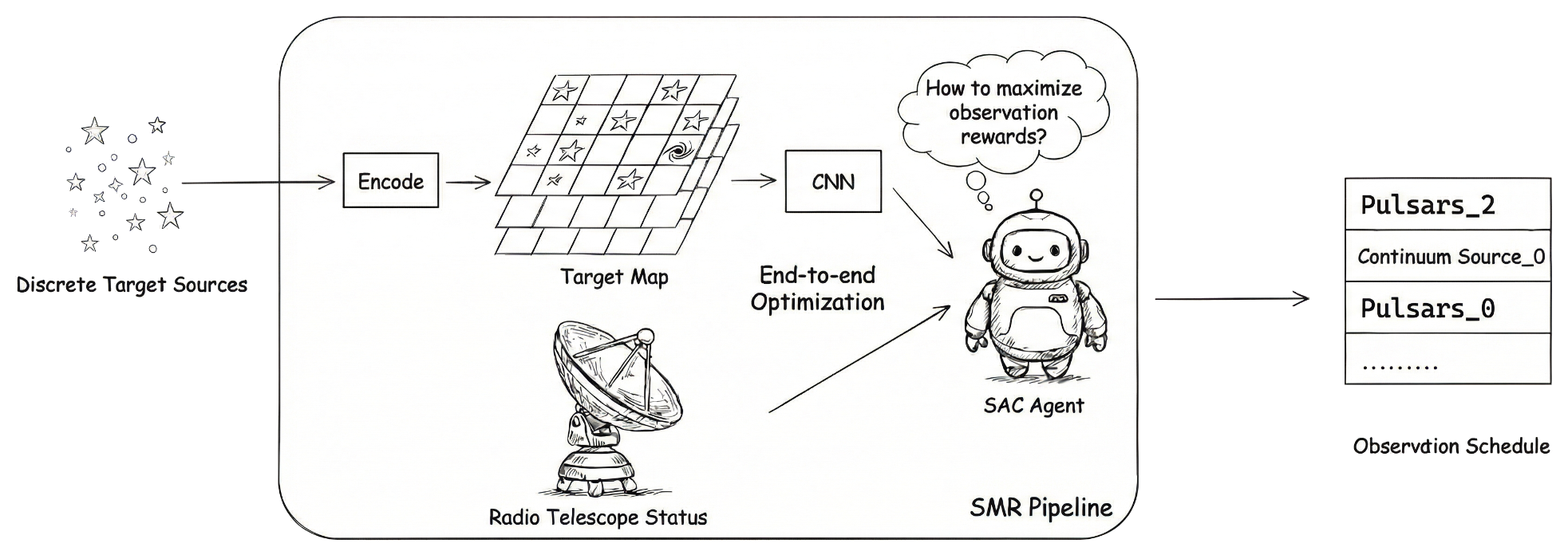}
    \caption{End-to-end framework of SMR.
    Discrete target attributes are projected to an Az--El horizon grid and encoded by a CNN, whose output is concatenated with the telescope state and passed to the SAC.
    All learnable components inside the box (CNN, feature fusion, and SAC) are optimized jointly end-to-end via gradient-based learning under the RL objective; the environment is used only to generate trajectories and rewards.}
  \label{fig:framework}
\end{figure}

For completeness, Figure~\ref{fig:framework} sketches the end-to-end pipeline \citep{huang_2026_19343630}.

\subsection{Action parameterization and feasibility projection}\label{subsec:action}
Directly outputting a discrete target index couples the policy’s output space to the (time-varying) number of visible targets and complicates deployment across instances.
Instead, we adopt a continuous action parameterization followed by a feasibility-preserving projection.

The policy outputs a continuous prototype action
\begin{equation}
\tilde{\mathbf{a}}_t=\big[\tilde{Az}_t,\ \tilde{El}_t,\ \tilde{r}_t,\ \tilde{u}_t\big],
\end{equation}
where $(\tilde{El}_t,\tilde{Az}_t)$ represent a preferred pointing and $(\tilde{r}_t,\tilde{u}_t)$ represent a preferred configuration.

At time $t$, the feasible target set is
\begin{equation}
\mathcal{A}_t=\{\,i: W_i(t)=1 \}.
\end{equation}
We project $\tilde{\mathbf{a}}_t$ to a discrete feasible target $i^*\in\mathcal{A}_t$ using a nearest-neighbor rule
in the joint space of $(El,Az,r,u)$ (with optional scaling coefficients if needed).
This guarantees that each executed decision is feasible while allowing the policy to remain continuous and fixed-dimensional.

\subsection{Transition dynamics, constraints, and reward}\label{subsec:dynamics}
The simulator evolves telescope states by explicitly modeling slew motion, a unidirectional azimuth wrap constraint,
instrument switching overhead, and Sun avoidance. These effects are reflected both in the state transition and in the
instantaneous reward.

We assume constant-speed motion on each axis. Let $\Delta El$ be the elevation change and $|\Delta Az|_{\mathrm{wrap}}$ be the
minimal wrapped azimuth change. The slew time is defined as the maximum of the two axis times,
\begin{equation}\label{Tslew}
T_{\mathrm{slew}}(t\!\rightarrow\!i)=
\max\!\Big\{\frac{|\Delta Az|_{\mathrm{wrap}}}{v_{Az}},\ \frac{|\Delta El|}{v_{El}}\Big\},
\end{equation}
where $(v_{El},v_{Az})$ are axis speeds in deg\,min$^{-1}$. 
Here, $T_{\rm slew}$ represents the transition-time overhead between two consecutive pointings. We approximate this overhead using a constant angular speed, which provides a simple and reproducible estimate of the time required to reposition the telescope.

The azimuth axis is subject to a unidirectional accumulated rotation limit of $270^\circ$.
When the accumulated rotation exceeds this limit, a wrap recovery is triggered and incurs a fixed penalty
$\lambda_{\mathrm{th}}$.

Instrument configuration switching is assumed to occur in parallel. The switching time is
\begin{equation}
T_{\mathrm{switch}}(t\!\rightarrow\!i)=\max\{\,\tau_r(r_t\!\rightarrow\!r_i),\ \tau_u(u_t\!\rightarrow\!u_i)\,\},
\end{equation}

where $\tau_r(\cdot)$ and $\tau_u(\cdot)$ are lookup times determined by the receiver and backend types.

We penalize the selection of targets within grid cells occupied by the Sun. The term $T_{\odot}(t)$ represents the specific time associated with observing a target inside this exclusion region.

The instantaneous reward balances gain-weighted on-source time against operational overheads:
\begin{equation}
R_t=
\underbrace{\bigl(1+\hat g_{r_{i}}(El_{i}(t))\bigr)
\,\bigl(2-\lambda_{s}(t)\bigr)\,t_{\rm on}}_{\text{on-source utility}}
-\lambda_{\mathrm{slew}}\,T_{\mathrm{slew}}
-\lambda_{\mathrm{switch}}\,T_{\mathrm{switch}}
-\lambda_{\odot}\,T_{\odot}(t)
-\lambda_{\mathrm{th}}.
\end{equation}
Here $t_{\rm on}$ is the on-source time accrued at this step and $\hat g$ is the normalized receiver gain.
If the gain channel or satellite interference channel is disabled, we set $\hat g\,\lambda_{s}=0$.

\subsection{Policy learning and training protocol}\label{subsec:training}
We learn the scheduling policy with Soft Actor--Critic (SAC) \citep{SAC} using \textsc{Stable-Baselines3} \citep{stable-baselines3}.
Training is divided into two stages.
Stage~0 sets switching costs to zero and masks instrument features, focusing on learning sky motion, slew kinematics, and the azimuth wrap constraint. This provides a stable initialization for the full problem.
Stage~1 restores the full dynamics, including instrument configuration switching costs, and fine-tunes the policy starting from the Stage~0 weights.
Hyperparameters follow Table~\ref{tab:train_hparams}; unless specified, we use default SAC settings from \textsc{Stable-Baselines3}.

\begin{table}[htbp]
    \centering
    \caption{Training hyperparameters for SAC. Unless otherwise stated, we follow the \textsc{Stable-Baselines3} SAC defaults.}
    \label{tab:train_hparams}
    \begin{tabular}{|l|c|c|}
        \hline
        \textbf{Parameter} & \textbf{Stage 0 (pretrain)} & \textbf{Stage 1 (fine-tune)} \\
        \hline
        Axis speeds $(v_{El},\,v_{Az})$ (deg\,min$^{-1}$) & 32, 24 & 32, 24 \\
        \hline
        Replay buffer size & $10^{6}$ & $10^{6}$ \\
        \hline
        Batch size & 256 & 256 \\
        \hline
        Actor learning rate & $3\times10^{-4}$ & $1\times10^{-4}$ \\
        \hline
        Critic learning rate & $3\times10^{-4}$ & $1\times10^{-4}$ \\
        \hline
        Discount factor $\gamma$ & 0.99 & 0.99 \\
        \hline
        Target update coefficient $\tau$ & 0.005 & 0.0025 \\
        \hline
        Entropy coefficient mode & auto & auto \\
        \hline
        Wrap-trigger penalty $\lambda_{\mathrm{th}}$ & 20 & 20 \\
        \hline
        Penalty weights $(\lambda_{\mathrm{slew}},\,\lambda_{\mathrm{switch}},\,\lambda_\odot)$  & 1, 1, 5 & 1, 1, 5 \\
        \hline
    \end{tabular}
\end{table}



\subsection{Baselines, ablation models, and evaluation metrics}
\label{subsec:baseline}

In this section, we introduce the baseline and ablation models used to evaluate SMR, together with the evaluation metrics adopted in our experiments. We use two complementary reference methods: a tuned look-ahead greedy heuristic, denoted as Greedy, and a Multi-Layer Perceptron baseline, denoted as MLP. Greedy serves as a non-learning scheduling baseline, whereas MLP is used as an architectural ablation to isolate the contribution of the map-encoded representation.

Greedy operates on the same feasible target set $\mathcal{A}_t$ and uses the same transition model, reward function, slew cost, switching cost, Sun-avoidance penalty, and azimuth-wrap logic as SMR. At each decision step, Greedy enumerates all currently feasible targets and assigns each candidate a score based on its immediate reward and a short look-ahead estimate:
\begin{equation}
S_{\rm greedy}(i)
=
w_1 R(t\rightarrow i)
+
w_2
\frac{1}{|\mathcal{N}_k(i)|}
\sum_{j\in \mathcal{N}_k(i)}
R(t_i\rightarrow j),
\label{eq:greedy_score}
\end{equation}
where $R(t\rightarrow i)$ is the immediate reward of selecting target $i$, $t_i$ denotes the simulator state after this virtual selection, and $\mathcal{N}_k(i)$ is the set of up to $k=\lfloor 0.1N_{\rm day}\rfloor$ feasible targets considered in the look-ahead evaluation from state $t_i$. If no feasible follow-up target is available, the look-ahead term is set to zero. The weights $(w_1,w_2)$ are selected by grid search over $(w_1,w_2)\in\{0.1,\ldots,0.9\}^2$ to maximize total return. Therefore, Greedy should be interpreted as a tuned myopic-plus-look-ahead heuristic rather than as a nearest-target rule.

The second reference method is a MLP baseline. This baseline is designed to isolate the contribution of the map-encoded representation. It uses the same SAC learning algorithm, the same continuous action parameterization, and the same feasibility projection as SMR, but replaces the CNN-based Az--El map encoder with a fully connected network that receives a flattened vector of per-target features. In the target-only setting, this vector is constructed from
\begin{equation}
\left\{({\rm Az},{\rm El},r,u,d)_i\right\}_{i=1}^{N}.
\end{equation}
When satellite-interference or gain information is enabled, the corresponding per-target values are appended to the same vector input. Thus, MLP is given access to the same candidate-level information as SMR, and the comparison tests whether arranging this information as a sky-aligned map provides an advantage over a sequentialized vector representation. To make the architectural comparison fair, the MLP network is tuned to have a parameter count within 5\% of the CNN-based SMR policy network.

Table~\ref{tab:baseline_summary} summarizes the role of each method in the comparison.

\begin{table*}
\centering
\caption{Summary of the schedulers compared in this work. Greedy is used as the main heuristic baseline, whereas MLP is used as an architectural ablation to test the necessity of the map-encoded representation.}
\label{tab:baseline_summary}
\renewcommand{\arraystretch}{1.15}
\begin{tabular}{p{0.11\textwidth} p{0.12\textwidth} p{0.30\textwidth} p{0.25\textwidth} p{0.16\textwidth}}
\hline
Method & Learning & Input representation & Decision rule \\
\hline
Greedy
& No
& Current feasible target list, optionally augmented with satellite occupancy and gain
& Selects the target with the largest tuned immediate-plus-look-ahead score\\

MLP
& Yes
& Flattened vector of per-target features concatenated with the telescope state, optionally augmented with satellite occupancy and gain
& SAC outputs a continuous prototype action, followed by the same feasibility projection as SMR\\

SMR
& Yes
& CNN-encoded multi-channel Az--El map concatenated with the telescope state
& SAC outputs a continuous prototype action, followed by feasibility projection \\
\hline
\end{tabular}
\end{table*}

Our primary evaluation metric for all methods is the time utilization
\begin{equation}
{\rm Util} = \frac{T_{\rm on}}{T_{\rm total}},
\label{eq:util}
\end{equation}
where $T_{\rm on}$ is the total on-source time and $T_{\rm total}=H$ is the scheduling horizon. We additionally report the number of observed targets $N$ within the horizon.

\section{Results and Analysis} \label{sec:results}

Following the experimental setup and metrics defined above, we evaluate the performance of SMR across varying scheduling horizons. To test the scalability and stability of the proposed method, we constructed test sets with five different window lengths: 2, 4, 8, 12, and 24 hours, denoted as Blocks A through E, respectively. For each test set, we report the average metrics over 20 independent test episodes to ensure statistical reliability.
Note that for the comparative analysis in Sections \ref{subsec:stage0} and \ref{subsec:stage1}, SMR utilizes a base configuration where the map encoder receives only the first channel (Target Embedding). This allows us to evaluate the core path-planning capability before introducing auxiliary information like satellite interference and gain, which are analyzed in Section \ref{sec:multichannel_analysis}.

\subsection{Performance on Kinematic Constraints (Stage 0)}\label{subsec:stage0}

First, we assess the scheduler's ability to handle fundamental telescope kinematics in Stage 0, where instrument switching costs are ignored ($\lambda_{\mathrm{switch}}=0$). 
Table \ref{tab:stage0_results} summarizes the performance comparison. In short observation windows (e.g., Block A, 2\,h), both SMR and Greedy baseline achieve high utilization rates ($\sim$93\% vs 90\%). This is expected, as short-term greedy decisions are often sufficient when the horizon is limited. However, as the planning horizon extends, the complexity of the combinatorial optimization increases. In the 24-hour test cases (Block E), SMR maintains a high utilization of 0.92, whereas Greedy baseline drops to 0.88. This indicates that the RL scheduler learns a more efficient long-term strategy that better navigates the geometric constraints of the sky, rather than maximizing only immediate rewards.

\begin{table}[htbp]
    \centering
    \caption{Performance Comparison in Stage 0 (Kinematics Only). Comparison of average number of observed targets ($N$) and time utilization (Util) between SMR and Greedy strategies.}
    \label{tab:stage0_results}
    \begin{tabular}{lcccccc}
        \toprule
        \textbf{Test Set} & \textbf{Duration} & \multicolumn{2}{c}{\textbf{SMR (Ours)}} & \multicolumn{2}{c}{\textbf{Greedy}} & \textbf{Gain} \\
        \cmidrule(lr){3-4} \cmidrule(lr){5-6}
         & (hours) & $N$ & Util & $N$ & Util & $\Delta$Util \\
        \midrule
        Block A & 2  & 20.3 & 0.93 ($\pm$0.028) & 19.9 & 0.91 ($\pm$0.027) & +0.02 \\
        Block B & 4  & 41.0 & 0.94 ($\pm$0.025) & 39.7 & 0.91 ($\pm$0.025) & +0.03 \\
        Block C & 8  & 81.2 & 0.93 ($\pm$0.013) & 78.6 & 0.90 ($\pm$0.018) & +0.03 \\
        Block D & 12 & 121.7 & 0.93 ($\pm$0.013) & 115.2 & 0.88 ($\pm$0.017) & +0.05 \\
        Block E & 24 & 243.5 & 0.93 ($\pm$0.013) & 230.5 & 0.88 ($\pm$0.018) & +0.05 \\
        \bottomrule
    \end{tabular}
    \footnotesize
    \begin{flushleft}
    Note: Values report the mean over 20 independent test instances for each horizon length $H\in\{2,4,8,12,24\}$ hours (Blocks A--E). Standard deviations for Util are shown in parentheses.
    \end{flushleft}
\end{table}

\subsection{Full Optimization with Instrument Switching (Stage 1)}\label{subsec:stage1}

In Stage 1, we introduce realistic overheads for switching instrument configuration. This reflects the operational reality of large single-dish telescopes, where changing the instrument configuration can incur substantial costs.
Table \ref{tab:stage1_results} presents the results. The introduction of switching costs inevitably reduces the theoretical maximum utilization for both methods compared to Stage 0. However, the performance gap between SMR and Greedy becomes substantially more pronounced.

While the two methods perform comparably in the shortest window (2\,h), SMR demonstrates superior scalability. In the 12-hour and 24-hour windows, SMR achieves a utilization of 0.86, significantly outperforming Greedy algorithm's 0.78 ($\Delta$Util = 0.08, or about a 10\% relative improvement ). This improvement in telescope time efficiency translates to approximately 1.7 hours of additional scientific observation per day, a substantial gain for oversubscribed facilities.

\begin{table}[htbp]
    \centering
    \caption{Performance Comparison in Stage 1 (Full Switching Costs). SMR significantly outperforms Greedy baseline in long-duration windows.}
    \label{tab:stage1_results}
    \begin{tabular}{lcccccc}
        \toprule
        \textbf{Test Set} & \textbf{Duration} & \multicolumn{2}{c}{\textbf{SMR (Ours)}} & \multicolumn{2}{c}{\textbf{Greedy}} & \textbf{Gain} \\
        \cmidrule(lr){3-4} \cmidrule(lr){5-6}
         & (hours) & $N$ & Util & $N$ & Util & $\Delta$Util \\
        \midrule
        Block A & 2  & 16.8 & 0.77 ($\pm$0.028) & 16.9 & 0.76 ($\pm$0.028) & +0.01 \\
        Block B & 4  & 33.5 & 0.77 ($\pm$0.023) & 32.7 & 0.75 ($\pm$0.027) & +0.02 \\
        Block C & 8  & 73.4 & 0.84 ($\pm$0.019) & 68.1 & 0.78 ($\pm$0.036) & +0.06 \\
        Block D & 12 & 112.5 & 0.86 ($\pm$0.020) & 102.1 & 0.78 ($\pm$0.036) & +0.08 \\
        Block E & 24 & 225.2 & 0.86 ($\pm$0.020) & 204.2 & 0.78 ($\pm$0.037) & +0.08 \\
        \bottomrule
    \end{tabular}
    \footnotesize
    \begin{flushleft}
    Note: Stage 1 includes both slew times (with cable wrap logic) and parallel instrument switching costs $T_{\mathrm{switch}} = \max \{ T_{\rm rx}, T_{\rm bk} \}$. 
    Values report the mean over 20 independent test instances for each horizon length $H\in\{2,4,8,12,24\}$ hours (Blocks A--E). Standard deviations for Util are shown in parentheses.
    \end{flushleft}
\end{table}

\begin{figure}[htbp]
    \centering
    \includegraphics[width=1.0\textwidth]{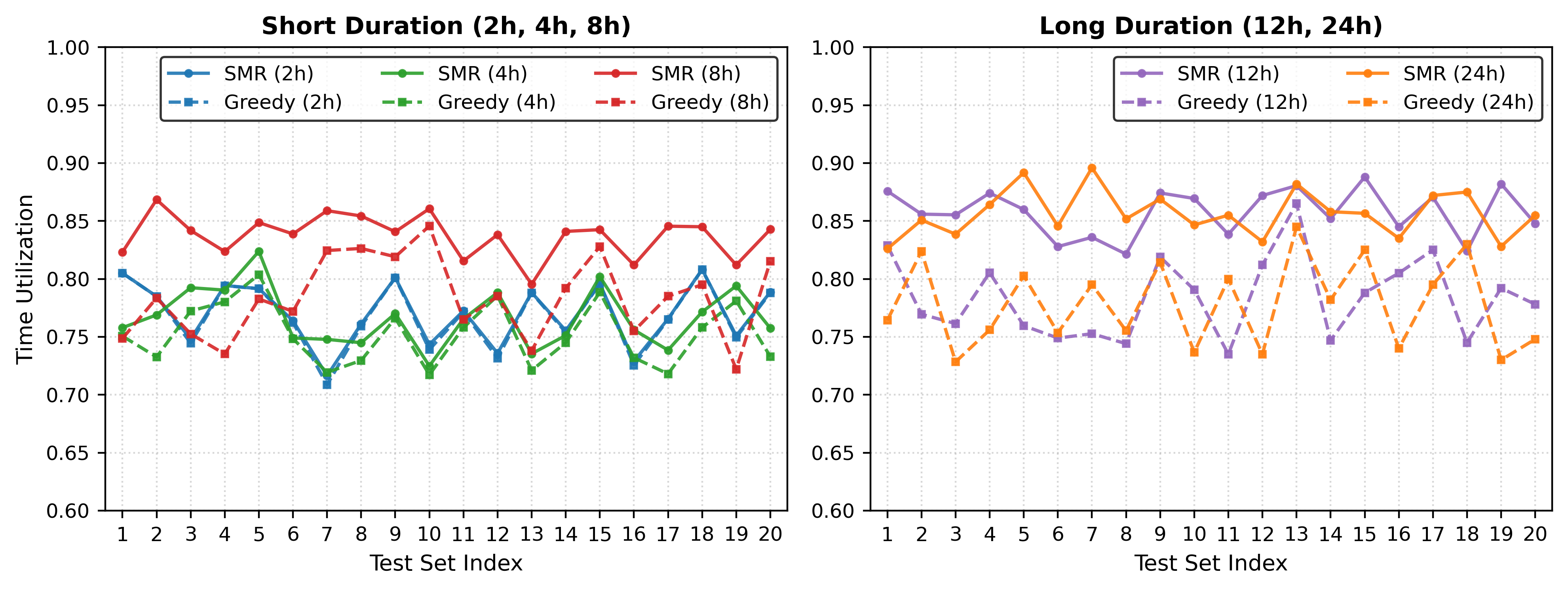}
    \caption{Utilization curves for SMR (solid) versus Greedy (dashed) across five time windows. Shaded regions indicate the standard deviation. SMR demonstrates greater stability and higher efficiency, particularly in 12-hour and 24-hour schedules.}
    \label{fig:util_curves}
\end{figure}

To visualize the stability of the proposed method, Figure \ref{fig:util_curves} plots the utilization trajectories for all test scenarios. SMR (solid lines) exhibits tighter confidence intervals and consistently higher mean performance across longer horizons compared to Greedy baseline (dashed lines), which tends to degrade as the scheduling window expands.

\subsection{Strategy Analysis: Balancing Slews and Switches} \label{subsec:strategy}

To understand the mechanism behind SMR's superior performance in long-duration schedules, we analyze the specific decision patterns regarding instrument switching and telescope slewing.

Table \ref{tab:switch_efficiency} compares the instrument switching frequency, quantified by the switching ratio $\rho$ (defined as the total number of switches divided by the number of unique instrument configurations in the target set). A value of $\rho=1.0$ indicates an ideal sequence with zero redundant instrument switching.

Counterintuitively, the results reveal that SMR performs more instrument switches than Greedy baseline, particularly in the 12-hour and 24-hour windows ($\Delta \rho \approx -0.16$, indicating SMR switches more frequently). In a naive short-term optimization, increasing switching overhead would typically reduce efficiency. However, as shown in the last column of Table \ref{tab:switch_efficiency}, the average cost of these additional switches ($\overline{\Delta t}_{(r, u)}$) is approximately 1.8 minutes.

\begin{table}[htbp]
    \centering
    \caption{Instrument Switching Analysis in Stage 1. $\rho$ represents the switching ratio (lower is fewer redundant switches). $\Delta \rho = \rho_{\text{SMR}} - \rho_{\text{Greedy}}$. A negative $\Delta \rho$ implies SMR switches more often. $\overline{\Delta t}_{(r, u)}$ denotes the average time cost of the additional switches incurred by SMR.}
    \label{tab:switch_efficiency}
    \begin{tabular}{lccccc}
        \toprule
        \textbf{Test Set} & \textbf{Duration} & \textbf{$\rho$ (SMR)} & \textbf{$\rho$ (Greedy)} & \textbf{$\Delta \rho$} & \textbf{$\overline{\Delta t}_{(r, u)}$ (min)} \\
        \midrule
        Block A & 2\,h  & 1.07 & 1.07 & 0.00  & 0.00 \\
        Block B & 4\,h  & 1.16 & 1.16 & 0.00  & 0.00 \\
        Block C & 8\,h  & 1.32 & 1.26 & -0.06 & 1.2 \\
        Block D & 12\,h & 1.36 & 1.26 & -0.10 & 1.8 \\
        Block E & 24\,h & 1.42 & 1.26 & -0.16 & 1.8 \\
        \bottomrule
    \end{tabular}
    \begin{flushleft}
    Note: Values report the mean over 20 independent test instances for each horizon length $H\in\{2,4,8,12,24\}$ hours (Blocks A--E). 
    \end{flushleft}
\end{table}

\begin{figure}[htbp]
    \centering
    \includegraphics[width=1.0\textwidth]{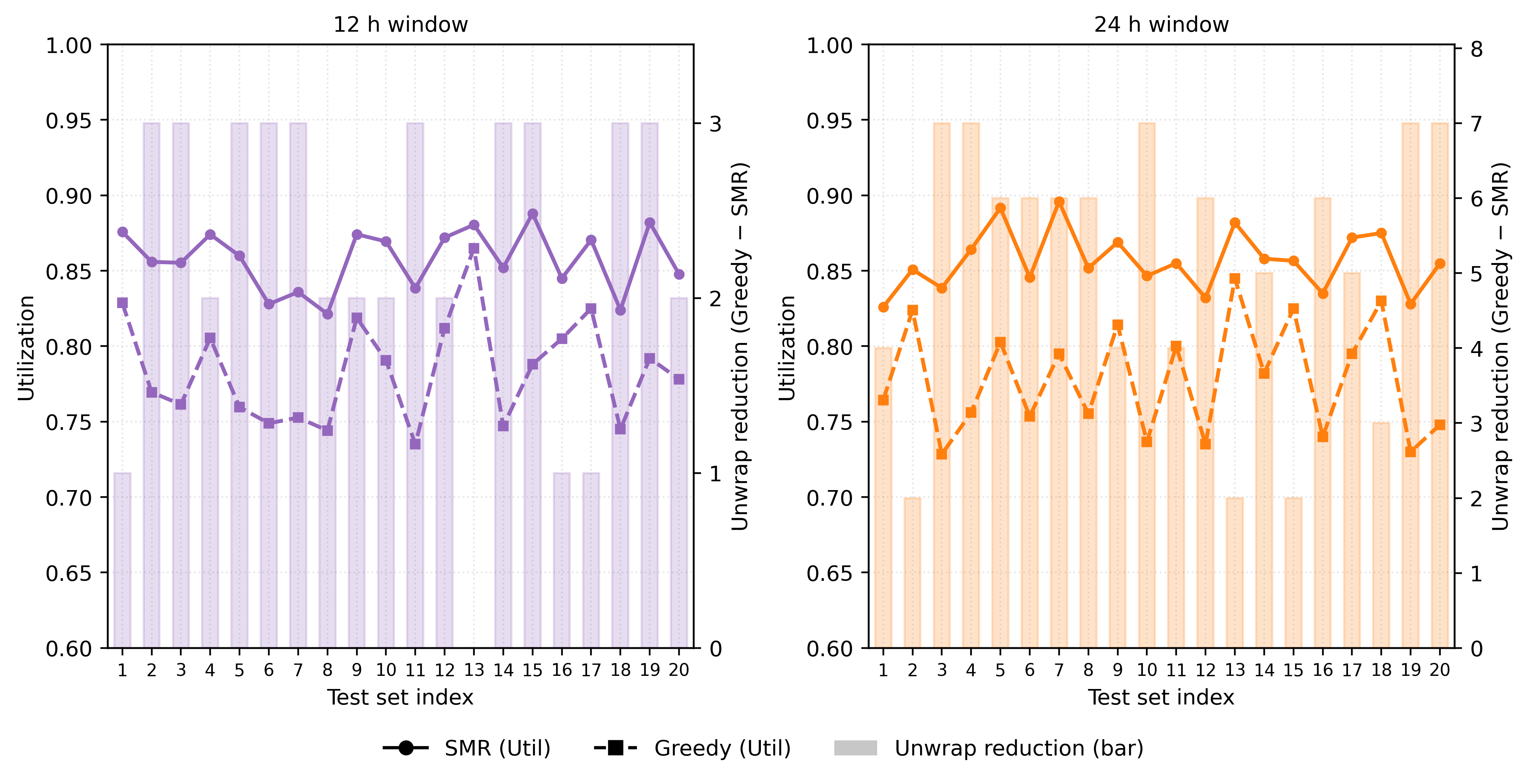}
    \caption{Reduction in cable unwrap events (Bars; Greedy - SMR) versus Efficiency Gain (Lines; Dual Axis) for 12h and 24h windows. Positive bars indicate that SMR triggered fewer cable unwraps than Greedy baseline. The scheduler effectively trades instrument switching costs for a reduction in time-consuming large-angle maneuvers.}
    \label{fig:wrap_reduce}
\end{figure}

This behavior reflects a trade-off between small switching costs and avoiding costly unwrap events.  Figure \ref{fig:wrap_reduce} illustrates the reduction in cable unwrap events achieved by SMR compared to Greedy in the 12-hour and 24-hour scenarios.
The data demonstrates a clear correlation: SMR consistently reduces or maintains the number of unwrap events across all test sets. By strategically accepting a short instrument switching penalty ($\sim$2 minutes), the scheduler avoids the much costlier unwrap maneuvers (which can take significantly longer and disrupt observation continuity). This behavior indicates that the RL scheduler has moved beyond greedy heuristics and learned a long-horizon planning strategy: it willingly incurs small, immediate costs to prevent catastrophic delays in the future.

\subsection{Ablation Study: The Necessity of Map-Encoding} \label{subsec:ablation_map}

A key innovation of SMR is the projection of target lists into a spatial image representation. To verify whether this explicit spatial encoding is necessary, we conducted an ablation study comparing SMR (using the Az--El Map) against a standard Deep Learning baseline using a Multi-Layer Perceptron (MLP). MLP baseline receives a flattened input vector of the target list $\left\{ (Az, El, r, u, d)_i \right\}_{i=1}^{N}$, ordered by index. To ensure a fair comparison, MLP architecture was tuned to have a parameter count within 5\% of SMR's CNN-based network. Both models were trained on the same 24-hour simulation datasets.

Table \ref{tab:ablation_map} details the performance gap between the two architectures. A striking trend emerges when analyzing the performance decay across different time windows.

\begin{table}[htbp]
    \centering
    \caption{Ablation Study: SMR (Az--El Map Input) vs. MLP (Vector Input). The $\Delta$ column represents the performance gain of SMR over MLP. While MLP performs adequately on the 24-hour task it was trained for, it fails to generalize to shorter windows.}
    \label{tab:ablation_map}
    \begin{tabular}{lcccc}
        \toprule
        \textbf{Test Set} & \textbf{Duration} & \textbf{Util (SMR)} & \textbf{Util (MLP)} & \textbf{Improvement ($\Delta$)} \\
        \midrule
        Block A & 2\,h  & 0.77 & 0.70 & +0.07 \\
        Block B & 4\,h  & 0.77 & 0.70 & +0.07 \\
        Block C & 8\,h  & 0.84 & 0.73 & +0.11 \\
        Block D & 12\,h & 0.86 & 0.74 & +0.12 \\
        Block E & 24\,h & 0.86 & 0.83 & +0.03 \\
        \bottomrule
    \end{tabular}
    \footnotesize
    \begin{flushleft}
    Note: Values report the mean over 20 independent test instances for each horizon length $H\in\{2,4,8,12,24\}$ hours (Blocks A--E).
    \end{flushleft}
\end{table}

On the 24-hour test set (Block E), which matches the duration of the training episodes, MLP achieves a respectable utilization of 0.83, only 0.03 behind SMR. However, its performance collapses on shorter windows (e.g., 8\,h and 12\,h), where the gap widens significantly ($\Delta > 0.1$).

This phenomenon suggests that the pure MLP architecture tends to memorize the global diurnal motion patterns or specific target trajectories present in the 24-hour training data. When the observation window is truncated or shifted (as in the shorter test blocks), this global memory fails. In contrast, SMR leverages the Spatial Inductive Bias provided by the Map-Encoder. By perceiving the density and geometry of targets on the local sky grid, the CNN learns a generalized policy based on the current state of the sky rather than memorized trajectories. This allows SMR to maintain robust performance regardless of the window length.

\section{Analysis of Multi-Channel Map Scalability} \label{sec:multichannel_analysis}

A key advantage of SMR is that additional constraints can be incorporated as extra channels without changing the action interface.
Unlike vector-based representations that often necessitate network restructuring to accommodate new features, SMR supports the seamless integration of diverse physical constraints as additional image channels. In this section, we analyze how extending the map depth enables the scheduler to make more balanced decisions, effectively weighing quantity against interference, and efficiency against signal quality.

\subsection{Channel 2: Satellite-interference proxy}

Satellite constellations can introduce direction-dependent RFI that varies with time.
In this channel, we use the satellite occupancy map (Channel~2) as a simple proxy for interference: observing in a direction with higher satellite density is assumed to be more prone to RFI.


To quantify whether the scheduler actually exploits Channel~2, we define the Low-Interference Exposure Ratio (LIER) as the fraction of on-source observing time spent in low-interference cells.
Using a fixed threshold $\lambda_{\mathrm{th}}=0.1$, we compute
\begin{equation}
\mathrm{LIER} =
\frac{\sum_{k} d_k \, \mathbf{1}\!\left[\lambda_{i_k}(t_k) \le \lambda_{\mathrm{th}}\right]}
{\sum_{k} d_k},
\end{equation}
where $k$ indexes the executed observations in an episode, $i_k$ is the selected target, $t_k$ is its decision time, and $d_k$ is its on-source integration time.
A higher LIER indicates that the scheduler allocates a larger fraction of observing time to directions with low satellite occupancy.

To isolate the benefit of the map-based spatial representation, we compare the 2-channel SMR (Target + Satellite) against an MLP and Greedy baselines.
For a fair comparison, MLP baseline is explicitly provided with the same real-time satellite occupancy value $\lambda_i(t)$ for every candidate target in its input vector: $\left\{ (Az, El, r, u, d, \lambda_i(t))_i \right\}_{i=1}^{N}$.
Thus, any performance gap reflects differences in representation and policy learning rather than access to privileged information.

Table~\ref{tab:sat_lier} reports the results.
SMR achieves a markedly higher LIER in both horizons (0.88 for 12\,h and 24\,h), indicating that it allocates substantially more on-source time to low-interference directions ($\lambda \le 0.1$).
Importantly, this improvement does not come at the cost of efficiency: SMR also attains higher utilization (0.83) than both MLP and Greedy baselines, while incurring only a marginal $\sim$0.03 drop in utilization compared to SMR without Channel~2.
As a validation check, a uniform-random scheduler typically attains LIER around $\sim$0.6 under the same threshold, further highlighting the significance of the achieved LIER=0.88.
Although our interference model is admittedly preliminary, introducing Channel~2 allows the agent to directly perceive a location-resolved interference signal.
As a result, SMR can exploit this spatial cue to proactively steer observations away from satellite-dense directions, learning an explicit avoidance behavior for interference sources rather than relying on indirect heuristics.

\begin{table}[htbp]
    \centering
    \caption{Mitigation of satellite interference (Channel~2). LIER measures the fraction of on-source time spent in low-occupancy directions ($\lambda \le 0.1$). A higher LIER indicates stronger avoidance of satellite-dense regions.}
    \label{tab:sat_lier}
    \begin{tabular}{lccccc}
        \toprule
        \textbf{Test Set} & \textbf{Duration} & \textbf{Method} & \textbf{Util} & \textbf{LIER} \\
        \midrule
        \multirow{3}{*}{Block D} & \multirow{3}{*}{12\,h}
            & SMR    & 0.83 & 0.88 \\
        &   & MLP    & 0.72 & 0.73 \\
        &   & Greedy & 0.73 & 0.72 \\
        \midrule
        \multirow{3}{*}{Block E} & \multirow{3}{*}{24\,h}
            & SMR    & 0.83 & 0.88 \\
        &   & MLP    & 0.79 & 0.78 \\
        &   & Greedy & 0.73 & 0.74 \\
        \bottomrule
    \end{tabular}
    \begin{flushleft}
    Note: Values report the mean over 20 independent test instances for each horizon length $H\in\{12,24\}$ hours (Blocks D, Blocks E).
    \end{flushleft}
\end{table}

\subsection{Channel 3: Gain}\label{gain}


Scientific yield depends not only on how much time the telescope spends on-source, but also on the instantaneous observation quality.
For single-dish radio observations, the signal-to-noise ratio is closely tied to the elevation-dependent gain curve, $\hat{g}(El)$.
With Channel~3 enabled, the scheduler is encouraged to favor high-gain pointings when such choices do not incur excessive overhead.

To quantify whether the learned policy indeed exploits gain information, we define the High Gain Observation Ratio (HGOR) as the fraction of on-source time spent in the high-gain regime:
\begin{equation}
\mathrm{HGOR}=\frac{\sum_k d_k\,\mathbf{1}\!\left[\hat{g}(t_k)\ge \hat{g}_{\mathrm{thresh}}(r_k)\right]}{\sum_k d_k},
\end{equation}
where $k$ indexes the executed observations in an episode, $d_k$ is the on-source integration time, and $\hat{g}(t_k)=\hat{g}(El_{i_k}(t_k))$ is the gain evaluated at the target's instantaneous elevation.
To ensure comparability across receivers with different gain curves, the threshold is defined per receiver as
\begin{equation}
    \hat{g}_{\mathrm{thresh}}(r) = \frac{4 \max (\hat{g}_{r}) + \min (\hat{g}_{r})}{5},
\end{equation}
which focuses on the high-gain regime near the upper end of each receiver's performance envelope.

To isolate the advantage of the map-based spatial representation, we compare the 2-channel SMR (Target + Gain) against MLP baseline and Greedy baseline.
MLP baseline is given access to the same gain information by augmenting its per-target input with the instantaneous gain,
$\left\{(Az,El,r,u,d,\hat{g}(t))_i\right\}_{i=1}^{N}$, following the same fixed-length target-vector construction used in the Target-only setting.
Therefore, any performance gap reflects differences in representation and policy learning rather than privileged information.

Table~\ref{tab:gain_quality} reports the results.
Although MLP baseline has access to the same per-target gain values as SMR, it still attains a substantially lower HGOR.
This suggests that encoding gain in a map-based representation with explicit direction information enables the agent to exploit the information more efficiently, resulting in a more effective allocation of on-source time to the high-gain regime.

\begin{table}[htbp]
    \centering
    \caption{Optimization of Observation Quality (Channel 3). SMR successfully trades a marginal amount of raw utilization for a significant boost in data quality (HGOR).}
    \label{tab:gain_quality}
    \begin{tabular}{lccccc}
        \toprule
        \textbf{Test Set} & \textbf{Duration} & \textbf{Method} & \textbf{Util} & \textbf{HGOR} \\
        \midrule
        \multirow{3}{*}{Block D} & \multirow{3}{*}{12\,h} 
            & SMR   & 0.82 & 0.42 \\
        &   & MLP   & 0.72 & 0.30 \\
        &   & Greedy& 0.75 & 0.29 \\
        \midrule
        \multirow{3}{*}{Block E} & \multirow{3}{*}{24\,h} 
            & SMR   & 0.83 & 0.42 \\
        &   & MLP   & 0.78 & 0.38 \\
        &   & Greedy& 0.76 & 0.30 \\
        \bottomrule
    \end{tabular}
    \begin{flushleft}
    Note: Values report the mean over 20 independent test instances for each horizon length $H\in\{12,24\}$ hours (Blocks D, Blocks E).
    \end{flushleft}
\end{table}

\subsection{Three-Channel SMR: Joint Optimization}

After isolating the effects of Channel~2 (satellite interference) and Channel~3 (gain) in the previous sections,
we now evaluate the full three-channel model, where the policy jointly observes the target map, the satellite-occupancy map, and the gain map.
The goal is to test whether the scheduler can simultaneously (i) avoid high-interference directions and (ii) favor high-gain observations,
while maintaining high time utilization.

We compare the three-channel SMR against MLP and Greedy baselines.
To ensure a fair comparison, MLP is provided with the same per-target satellite-occupancy and gain values,
by augmenting its input vector as $\{(Az,El,r,u,d,\lambda(t),\hat{g}(t))_i\}_{i=1}^{N}$ using the same fixed-length target-vector construction as in the previous baselines.
LIER is computed with the threshold $\lambda \le 0.1$, and HGOR uses the receiver-dependent threshold defined in Section~\ref{gain}.

Table~\ref{tab:three_channel} summarizes the three-channel setting, where the policy must jointly balance time utilization, satellite-interference avoidance, and high-gain observing.
Because the objectives are now coupled, LIER and HGOR are expected to be lower than in the corresponding single-channel settings.
Nevertheless, SMR achieves the best overall trade-off across both horizons: it attains the highest utilization (0.80), the highest LIER (0.82/0.81), and the highest HGOR (0.37/0.38).
Compared to MLP baseline, SMR improves LIER by +0.12 (12\,h) and +0.05 (24\,h), and improves HGOR by +0.13 (12\,h) and +0.06 (24\,h), while also maintaining higher utilization.

To put these results in context, we compare SMR against Greedy and MLP baseline.
Greedy remains consistently weaker on all three metrics, indicating that myopic selection is insufficient when interference and gain cues are both time-varying and spatially structured.
MLP baseline shows a different limitation: its performance is noticeably more sensitive to the evaluation horizon.
On the 24-hour test set, which matches the training horizon, MLP consistently outperforms Greedy baseline, suggesting that trial-and-error training in the RL framework can learn a more balanced policy than a purely myopic rule even with a vectorized input.
However, on the 12-hour test set, MLP degrades substantially and even falls below Greedy.
A plausible explanation is that, because target coordinates and visibility change continuously with time, a sequentialized fixed-length vector representation provides a weak inductive bias for capturing location-dependent structure.
This limitation becomes more pronounced as additional spatially varying signals are introduced (e.g., satellite occupancy and gain), since the policy must learn to fuse multiple fields that are inherently defined over pointing directions.
As a result, MLP baseline becomes more sensitive to the training horizon and less robust under horizon shifts.
In contrast, SMR encodes targets and auxiliary cues as aligned multi-channel Az--El maps, allowing the policy to exploit local spatial structure and integrate the added channels more directly, which leads to consistently stronger performance across horizons.

\begin{table}[htbp]
    \centering
    \caption{Three-channel scheduling performance (Target + Satellite + Gain). The full SMR jointly optimizes efficiency (Util), interference avoidance (LIER), and observation quality (HGOR).}
    \label{tab:three_channel}
    \begin{tabular}{lccccc}
        \toprule
        \textbf{Test Set} & \textbf{Duration} & \textbf{Method} & \textbf{Util} & \textbf{LIER} & \textbf{HGOR} \\
        \midrule
        \multirow{3}{*}{Block D} & \multirow{3}{*}{12\,h}
            & SMR     & 0.80 & 0.82 & 0.37 \\
        &   & MLP     & 0.70 & 0.70 & 0.24 \\
        &   & Greedy        & 0.75 & 0.73 & 0.27 \\
        \midrule
        \multirow{3}{*}{Block E} & \multirow{3}{*}{24\,h}
            & SMR     & 0.80 & 0.81 & 0.38 \\
        &   & MLP     & 0.76 & 0.76 & 0.32 \\
        &   & Greedy        & 0.75 & 0.73 & 0.28 \\
        \bottomrule
    \end{tabular}
    \begin{flushleft}
    Note: Values report the mean over 20 independent test instances for each horizon length $H\in\{12,24\}$ hours (Blocks D, Blocks E).
    \end{flushleft}
\end{table}

\section{Discussion}\label{sec:discussion}

\subsection{Extensible multi-channel sky representation}\label{subsec:disc_extensible}

A central contribution of SMR is the map-encoded observation.
It turns a variable-length target list into a fixed-size tensor on the local Az--El grid and can be trained end-to-end from the encoder to the policy.
This design is also easy to extend: new observing conditions can be added as extra map channels on the same grid, without changing the action space or the training setup.
In addition to adding new physical channels, the spatial discretization and the within-pixel target representation are also implementation choices rather than fixed assumptions of the framework. For denser or more clustered observing programs, one may use a finer Az--El grid or replace the single-occupancy target channel with an aggregation scheme that combines multiple target embeddings within the same pixel.

Our results illustrate why this matters.
When we add spatial signals such as the satellite-occupancy map (Channel~2) and the gain map (Channel~3), SMR can directly use these location-resolved cues and learn to avoid unfavorable directions while still seeking high-gain observations.
In contrast, a vector-based MLP baseline, even when given the same per-target values, is less effective at combining multiple sky-defined fields into a stable decision rule.

This extension mechanism matches how modern observatories operate.
Queue observing and dynamic scheduling are routinely used to adapt to time-varying conditions, and to re-prioritize tasks when conditions differ from what was planned \citep{Nyman2010ALMAScienceOps}.
From this viewpoint, Channel~2 and Channel~3 are just two examples: they encode interference risk and instrumental response in a sky-aligned form that the policy can reason about directly.

Looking forward, the same idea can incorporate other telescope-specific factors that affect feasibility and data quality.
Examples include atmospheric proxies (e.g., opacity or precipitable water vapor, and, where relevant, cloud cover) for forecast-aware scheduling \citep{YeChen2013CloudForecast,Nyman2010ALMAScienceOps},
and direction- and time-dependent interference indicators, motivated by the long-standing importance of RFI monitoring and mitigation in radio astronomy \citep{FridmanBaan2001RFIReview}.
Overall, SMR offers a simple interface for adding such factors as additional map channels, while encouraging the policy to use local sky structure.

\subsection{Sensitivity of SMR}\label{subsec:disc_reward}

Our reward is designed to trade on-source utility against operational overheads.
In practice, the relative scale of the penalty and weighting terms largely determines the scheduler's behavior, and several parameters are particularly influential.

The Sun-related penalty (set by $\lambda_\odot$ together with the exclusion scale) is easy to interpret relative to the maximum achievable per-step on-source utility.
Once $\lambda_\odot$ becomes comparable to (or larger than) the peak gain-weighted observing reward, the policy quickly learns to avoid Sun-proximate directions during slews and target transitions.
The wrap-trigger penalty $\lambda_{\mathrm{th}}$ shows a non-monotonic effect.
If it is too large, the scheduler can become overly conservative, preferring local moves or extra configuration changes to reduce wrap risk, at the cost of global efficiency.
If it is too small, wrap events become frequent and the policy drifts toward short-horizon, greedy behavior.
This suggests that $\lambda_{\mathrm{th}}$ is facility-dependent and should be calibrated to the specific mount geometry and operational tolerance, rather than treated as a universal constant.
By comparison, the switching penalties are more robust in our experiments: even when the effective switching time costs are doubled, performance degrades only mildly.
A natural interpretation is that the optimization target is the day-level cumulative return, so the scheduler can learn to amortize switching overheads by exploiting longer-term structure in the target distribution.

Conceptually, the gain channel and the satellite-interference channel play the same role: they introduce a location-dependent positive preference (favor high-gain directions) and a location-dependent negative preference (avoid interference-dense directions).
In our experiments, the corresponding reward weights are among the most sensitive parameters, especially when the values are not naturally normalized.
In that case, small changes in scaling can noticeably change the learned trade-off between efficiency, quality, and avoidance.

This observation also motivates a practical design choice.
When multiple sources of positive/negative preferences are present (e.g., gain, atmospheric quality, different classes of RFI, or project-specific priorities), encoding them as different separate map channels can be preferable to collapsing them into a single composite map.
A single-channel fusion typically requires normalization, which may blur relative contrasts, smooth out gradients, and make distinct effects harder for the policy to disentangle.
Multi-channel maps preserve the provenance of each factor and allow the network to learn how to combine them in a task-dependent way.

In addition to the policy and reward design, we tested how sensitive the results are to several choices in the simulator and in the construction of daily observing programs.
Overall, the scheduler is only weakly affected by the observing site and by the coarse mix of target classes in the catalog.
This is expected for our setting, because the dominant losses are set by telescope operations---mount slews, configuration changes, and cable-wrap management---rather than by detailed astrophysical properties of individual sources.

By contrast, the assumed distribution of per-target on-source times can have a noticeable impact.
Shorter typical integrations favor frequent retargeting and make transition overheads more important, whereas longer integrations reduce the relative cost of moving and shift the optimal balance toward fewer, higher-commitment observations.
As a result, changing the on-source time distribution effectively changes the ``duty cycle'' of the program and can require re-tuning the reward scaling to preserve the intended trade-off between efficiency and quality.

Finally, the wrap-trigger setting remains influential.
Unlike local slew costs, the wrap constraint is global and history-dependent: it depends on accumulated azimuth motion and can force costly corrective actions once the limit is reached.
Varying the trigger threshold therefore changes the long-horizon planning problem itself, and can lead the scheduler to adopt qualitatively different strategies for managing risk across the night.

Overall, these results reinforce that SMR---like most RL-based schedulers---is a coupled design of simulator and reward, rather than a general network architecture.
The most transferable component is the multi-channel sky-map interface: it provides a general way to present time-varying, direction-dependent operational context to the policy.
By contrast, the specific reward weights and numeric scales are not meant to be universal.
Future deployments should tune them to the telescope system, science goals, and observing preferences, and may benefit from exploring more realistic value settings and trade-offs as part of the commissioning workflow.


\subsection{Limitations and outlook}\label{subsec:disc_limitations}

Several limitations point to clear next steps.
First, some of our auxiliary channels are intentionally simplified and should be viewed as proof-of-concept inputs for testing extensibility rather than as site-ready operational models.
For example, the satellite interference map is built from a simplified occupancy proxy, and the gain channel uses an idealized elevation-dependent response curve.
Both choices are sufficient to validate the multi-channel interface, but they remain far from the full complexity of real RFI conditions and instrument calibration.
A natural extension is therefore to integrate forecast-informed and measurement-driven context, such as weather/opacity and time-varying RFI indicators, which the multi-channel design can accommodate without changing the action interface.

Second, the simulator adopts a simplified telescope motion and observing model.
Real facilities exhibit richer slew and settling behavior (e.g., acceleration limits, tracking and reacquisition transients, drive-dependent speed profiles, and mode-dependent overheads), whereas our current implementation assumes a simplified kinematic model and an idealized ``track-and-integrate'' observation once a target is selected.
Improving the realism of these dynamics---while preserving a tractable training environment---is important for narrowing the sim-to-operations gap.

Finally, while the present reward provides an effective single-scalar training signal, observatory operations are inherently multi-criteria.
Exposing explicit Pareto trade-offs (e.g., utilization versus data quality versus interference risk) and incorporating uncertainty-aware planning are promising directions for making SMR a more transparent decision-support tool in real scheduling pipelines.

\section{Conclusion}\label{sec:conclusion}

In this paper, we presented SMR (Scheduler with Map-encoded Reinforcement Learning), a novel framework designed to address the dynamic scheduling challenges of single-dish radio telescopes.
The central idea is a map-encoded observation interface: at each decision step, a variable-length target set and time-varying operational context are projected onto an Az--El grid as an aligned multi-channel sky map.
This converts a combinatorial, set-structured scheduling problem into a spatial representation that can be processed efficiently by convolutional policies, while preserving local sky geometry and avoiding ad hoc padding schemes as the candidate set changes.

We demonstrated that this representation leads to consistently stronger long-horizon scheduling performance than vector-based baselines.
In full-day (24-hour) simulations with mount kinematics, instrument switching overhead, Sun avoidance, and an azimuth wrap constraint, SMR achieves higher time utilization than a tuned look-ahead greedy heuristic (about a 10\% improvement in our main setting) and learns non-myopic behavior, such as accepting small switching overheads to reduce the risk of costly wrap events later in the schedule.
Ablation results further show that the map representation is a key contributor to robustness across different horizon lengths, whereas vectorized MLP policies are more sensitive to horizon shifts when the sky configuration changes over time.

Beyond efficiency, SMR provides a simple and extensible mechanism to incorporate additional observing preferences as new map channels.
Using a satellite-occupancy interference map (Channel~2) and an elevation-dependent gain map (Channel~3) as two concrete examples, we showed that SMR can leverage location-resolved cues to improve both interference avoidance and observation quality.
In the three-channel setting, SMR delivers the strongest joint trade-off across both horizons.
Relative to MLP baseline, SMR improves utilization by $\sim$14\% (12\,h) and $\sim$5\% (24\,h), increases LIER by $\sim$17\% and $\sim$7\%, and boosts HGOR by $\sim$54\% and $\sim$19\%, respectively.
SMR also remains consistently better than greedy baseline (e.g., $\sim$7\% higher utilization and $\sim$11--12\% higher LIER, with $\sim$36--37\% higher HGOR).
These results support the main conclusion of this paper: the multi-channel sky-map design is the most transferable component of SMR, enabling the policy to combine heterogeneous, direction-dependent factors in a geometrically aligned form.

Several limitations point to clear next steps.
Our current channels are simplified (e.g., a proxy satellite-density field and idealized gain curves) and should be viewed as proof-of-concept inputs for testing extensibility rather than site-ready operational models.
Bridging this gap will require replacing these proxies with measurement-driven and forecast-informed products, such as weather/opacity indicators and time-varying RFI monitors, which can be added as further map channels without changing the action interface.
More broadly, observatory scheduling is inherently multi-criteria; future work should expose explicit trade-offs (e.g., utilization versus quality versus interference risk) through multi-objective or constrained RL formulations and incorporate uncertainty-aware planning.
Finally, transfer and continual learning are promising directions: the map-based interface provides a natural common representation across telescopes and observing programs, offering a practical path to adapt simulation-trained policies to on-sky operations with minimal redesign.

\begin{acknowledgments}
This work was supported by National Key R\&D Program of China No.2021YFC2203501. The research was also partly supported by the Operation, Maintenance and Upgrading Fund for Astronomical Telescopes and Facility Instruments, budgeted from the Ministry of Finance of China (MOF) and administrated by the Chinese Academy of Sciences (CAS), and the Scientific Instrument Developing Project of the Chinese Academy of Sciences, Grant No. PTYQ2022YZZD01.

We would also like to thank Jingyi Deng, Haihao Shi, Junda Zhou, Xuwei Zhang and Yangyu Liu for helpful discussions on topics related to this work.
\end{acknowledgments}

\software{NumPy \citep{harris2020array},
          Astropy \citep{astropy1,astropy2,astropy3},
          SciPy \citep{2020SciPy-NMeth},
          Matplotlib \citep{Hunter:2007},
          PyTorch \citep{paszke2019pytorchimperativestylehighperformance},  
          scikit-learn \citep{scikit-learn}
          }

\newpage

\bibliography{sample701}{}
\bibliographystyle{aasjournalv7}



\end{document}